\documentclass[prl,aps,showpacs,twocolumn,superscriptaddress]{revtex4}

\newcommand{\Es}{\hat{E}}

\newcommand{\es}{\hat{{\cal E}}}

\newcommand{\rb}{{\bf r}}
\newcommand{\eb}{{\bf e}}
\newcommand{\rbt}{({\bf r},t)}
\newcommand{\rbpt}{({\bf r}^\prime,t)}
\newcommand{\kb}{{\bf k}}

\newcommand{\bra}[1]{{\langle #1|}}
\newcommand{\ket}[1]{{| #1\rangle}}
\newcommand{\braket}[1]{{\langle #1\rangle}}

\newcommand{\ds}{\displaystyle}
\newcommand{\dd}{\partial}
\newcommand{\dt}{\ds\frac{\dd}{\dd t}}
\newcommand{\dz}{\ds\frac{\dd}{\dd z}}
\newcommand{\dzz}{\ds\frac{\dd^2}{\dd z^2}}

\newcommand\be{\begin{eqnarray}}
\newcommand\ee{\end{eqnarray}}

\usepackage{graphicx}
\usepackage{epsfig}
\usepackage{amssymb}
\usepackage{amsmath}
\usepackage{amsfonts}
\usepackage{color}

\begin{document}
\title{Dipolar Bose-Einstein condensate of Stationary-Light Dark-state Polaritons}
\author{F. E. Zimmer}
\affiliation{Max Planck Institute for the Physics of Complex Systems, 01187 Dresden, Germany}
\author{G. Nikoghosyan}
\affiliation{Institut f\"ur Theoretische Physik, Albert-Einstein Allee 11, Universit\"at Ulm, 89069 Ulm, Germany}
\affiliation{Institute of Physical Research, 378410, Ashtarak-2, Armenia}
\author{M. B. Plenio}
\affiliation{Institut f\"ur Theoretische Physik, Albert-Einstein Allee 11, Universit\"at Ulm, 89069 Ulm, Germany}
\date{\today }

\begin{abstract}
We put forward and discuss in detail a scheme to achieve Bose-Einstein condensation of stationary-light dark-state polaritons with dipolar interaction. To this end we have introduced a diamond-like coupling scheme in a vapor of Rydberg atoms under the frozen gas approximation. To determine the system's dynamics we employ normal modes and identify the dark-state polariton corresponding to one of the modes. We show that in contrast to atomic dipolar ultra-cold vapors dark-state polariton Bose-Einstein condensates proposed here can be stable for a negative dipolar interaction constant.
\end{abstract}

\pacs{42.50.Gy, 03.75.Hh, 32.80.Rm}
\maketitle
%
%
Long before the breakthrough  of Bose-Einstein condensation (BEC) in atomic systems quasiparticles have been considered as promising candidates for condensation at high temperatures \cite{Blatt-PR-1962}. Their high critical temperature, which is one of their advantages compared to the atomic approach, is due to its inverse proportionality to the constituents' mass.
Despite early attempts the condensation of quasiparticles in form of micro-cavity polaritons and thin-film magnon has only recently been achieved \cite{Balili-Science-2007,Demokritov-Nature-2006}. These systems show, however, two major downsides: (a) a relative short lifetime of the quasiparticles with respect to the thermalization time and (b) the two-dimensionality of both systems. In particular (b) prevents  the formation of true long-range order of the one-particle density matrix and only quasicondensation can  be achieved.

The lifetime of quasiparticles can be extended if one uses, as shown in \cite{Fleischhauer-PRL-2008}, yet an other quasiparticle: the stationary-light dark-state polariton (SL-DSP) \cite{Zimmer-PRA-2008} a quantum superposition of photons and collective Raman excitations. Thermalization can be achieved by introducing a Kerr nonlinearity in the medium which effectively leads to a contact interaction of SL-DSP \cite{Fleischhauer-PRL-2008}.
Here we consider a different mechanism of interaction based on an optically thick ensemble of Rydberg atoms. The Rydberg atoms induce a non-local long-range interaction between SL-DSP and hence introduce a novel interaction between polaritonic quasiparticles not known from earlier approaches. This can be used to simulate the dynamics of dipolar ultra-cold atomic systems. Moreover we will show that by varying the orientation of the dipoles one can control the interaction potential and realize new many body phenomena, e.g. very stable 3D BEC of DSP with long-range interaction or break the radial symmetry of dipole-dipole interaction.

%
\begin{figure}[b]
 \begin{center}
    \includegraphics[width=9cm]{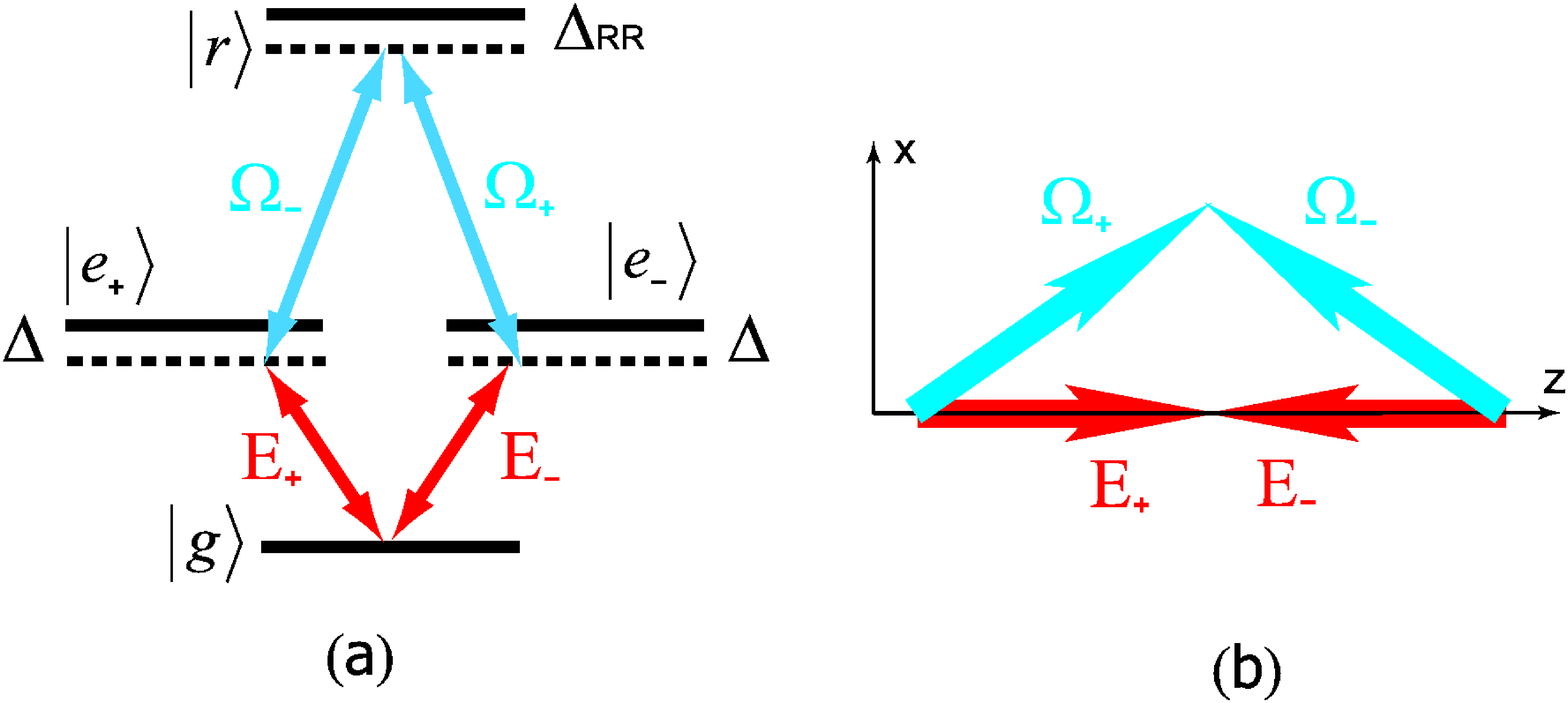}
  \caption{(Color online) The diamond linkage pattern (a) shown together with (b) the used field configuration for the counter-propagating control ($\Omega_\pm$) and induced probe fields ($\hat{E}_\pm$). $|g\rangle$ denotes ground and $|r\rangle$ Rydberg state. $|e_\pm\rangle$ are intermediate states. The one-photon detuning of the probe field transitions is denoted by $\Delta$. The dipolar interaction of the Rydberg atoms induces an effective two-photon detuning $\Delta_{RR}$. It is assumed that  $\omega_c\neq\omega$ and thus there is no cross-couplings between the probe and coupling fields. Phase-matching is achieved by using non-parallel probe and coupling fields.
\label{fig:diamond-linkage-pattern-and-setup}}
 \end{center}
\end{figure}
%
Let us consider a medium consisting of an ensemble of 4-level atoms with a diamond configuration resonantly interacting with four fields (Fig. \ref{fig:diamond-linkage-pattern-and-setup}).  The weak probe fields $\Es_\pm$ couple to the transitions $\ket{g}\leftrightarrow\ket{e_\pm}$ and strong control fields $\Omega_\pm$ drive the transitions $\ket{e_\mp}\leftrightarrow\ket{r}$. If laser fields are present in only one leg of the diamond level configuration, e.g. $\Es_+ = \Omega_- = 0$,  the scheme reduces to a 3-level ladder-configuration which is known to support electromagnetically induced transparency (EIT) \cite{Fleischhauer-RMP-2005}. This prominent effect has also been observed in Rydberg systems \cite{Mohapatra-PRL-2007,Friedler-PRA-2005}. Alongside transparency  a probe field experiences at two-photon resonance a dramatic change of the refractive index properties of the medium leading for instance to a slow group velocity of a probe field pulse \cite{Hau-Nature-1999}.

In the case that counter-propagating coupling fields in both legs are switched on and have comparable intensities the probe fields form a quasi-stationary pattern usually termed stationary light \cite{Bajcsy-Nature-2003,Andre-PRL-2005,Zimmer-OC-2006,Zimmer-PRA-2008,Fleischhauer-PRL-2008}.
We here consider the dynamics of stationary probe fields  $\hat{E}_\pm$  in an ensemble of Rydberg atoms as of Fig. \ref{fig:diamond-linkage-pattern-and-setup}.
The propagation of the probe fields is determined by the external control fields $\Omega_c^{(\pm)}$ which we assume to be sufficiently strong such that the medium shows no influence on them.
For the later discussion we introduce slowly varying (normalized) probe field amplitudes $\es_\pm(\rb,t)$ using  $\Es_\pm(\rb,t)=\sqrt{\frac{\hbar\omega}{2\epsilon_0}}\es_\pm(\rb,t)\exp\left[-i\left(\omega t\mp kz\right)\right]$ and control field amplitudes $\Omega_\pm(t)$ via
$\Omega_c^{(\pm)}=\Omega_\pm(t) \exp\left[-i\left(\omega_c t - \kb_c \rb\right)\right]$. In addition we introduce collective spin-flip operator $\hat\sigma_{\mu\nu}=\sum_{j\in V(\rb)}\ket{\mu}_{jj} \bra{\nu}/N(\rb)$ where $\ket{\mu}_j,\,\ket{\nu}_j$ denote single atom eigenstates and the averaging is taken over a small volume $V(\rb)$ around the position $\rb$ with $N(\rb)$ atoms inside \cite{Fleischhauer-PRA-2002}.

The scheme shown in Fig. \ref{fig:diamond-linkage-pattern-and-setup} as well as other linkage patterns \cite{Zimmer-PRA-2008} suffer from a coupling of the probe $\es_\pm$ (control $\Omega_\pm$) field to the transition $\ket{e_\mp}\leftrightarrow\ket{r}$ ($\ket{g}\leftrightarrow\ket{e_\pm}$) if the transition frequencies of the lower and upper legs of the diamond structures are comparable. In one dimensional systems discussed so far \cite{Bajcsy-Nature-2003,Zimmer-OC-2006,Zimmer-PRA-2008}, comparable transition frequencies have been employed to ensure EIT in hot atomic vapors with Doppler broadening \cite{Fleischhauer-RMP-2005}.  However, in order to avoid this cross-coupling we assume that frequencies of probe and coupling fields are not equal, i.e. $\omega_c\neq\omega$. This can be achieved by using the setup shown in Fig. 1 (b).

We note that an efficient 4-wave mixing process can only be achieved if the setup used fulfills a phase-matching condition. If we denote the wave vector component of the control field parallel to the propagation direction of the probe field by $\kb^\parallel_{c,\pm}$ and the perpendicular component by $\kb^\bot_{c,\pm}$ it is easy to show that the condition $\Delta \kb = \kb_++\kb_{c,-}-(\kb_-+\kb_{c,+})=0$ can be satisfied if one chooses $\kb^\parallel_{c,\pm}=-\kb_\mp=\pm k \eb_z$ and hence $\kb^\bot_{c,+}=\kb^\bot_{c,-}$.
Due to the non-vanishing perpendicular wave vector components of the control fields, we are restricted to the regime of cold Rydberg vapors.
The choice above also eliminates the decay of stationary-light \cite{Kiffner-PRA-2010}.

The light-matter interaction Hamiltonian including dipole-dipole interaction reads after rotating wave approximation and after eliminating  spatially fast changing contributions in the direction of propagation of the probe fields
\begin{align}
\hat{H}&=\hbar \frac{N}{V_t}\int_{V_t}\Big[\Delta \sum_{j=\pm}\hat\sigma_{e_je_j}\nonumber\\
&
-\left(\sum_{j=\pm}g\hat\sigma_{e_j g}\es_j\rbt
+
\hat\sigma_{r e_{-j}}\Omega_j(t) e^{ -i\kb_c^\bot \rb}+h.a.\right)
\label{eq:Hamiltonian}\\
&+\left.\frac{N}{2V_t}\int_{V_t}\hat\sigma_{rr}\rbt\hat\sigma_{rr}\rbpt\varepsilon(\rb^\prime-\rb){\rm d}^3 r^\prime\right]{\rm d}^3 r\nonumber
\end{align}
where $N$ is the number of atoms, $V_t$ denotes the interaction volume, and $\Delta$ is the one-photon detuning of $\es_\pm$ and
$
\varepsilon \left(\rb\right) =U\wp_{r}^2(1-3\cos ^{2}\varphi)\left\vert\rb\right\vert ^{-3}
$
denotes the frequency shift of the Rydberg state due the dipolar interaction of the atoms.
We assume that the dipole moments  $\wp_r$ of the Rydberg atoms are induced by a polarizing external field and thus are perfectly aligned \cite{Gallagher-1994}. The angle between the polarization direction of the dipoles and the
difference vector $\rb^\prime-\rb$ is denoted by $\varphi$.
It is well known that by using a rotating polarizing external field it is possible to tune the dipole-dipole interaction, i.e. to reduce its effective strength $U$ and to even change its sign \cite{Giovanazzi-PRL-2002,Lahaye-RPP-2009,Buechler-NP-2007}. This feature will be of importance later on.
In eq. (\ref{eq:Hamiltonian}) the atom-field coupling constants are given by $g=\wp\sqrt{\frac{\omega}{2\hbar\epsilon_0 V_t}}$ with
$\wp$ being the dipole moment of the $\ket{g}\leftrightarrow\ket{e_\pm}$-transitions. Finally we note that without Rydber-Rydberg interaction term the two photon detuning in each leg is zero.

In the following we describe the derivation of an effective equation of motion for the stationary-light dark-state polariton
corresponding to the considered diamond-like coupling scheme.
The propagation of the probe field is described by the Maxwell equation for the slowly varying amplitude in paraxial approximation%
\begin{equation}
 \left(\dt\pm c\dz-i\frac{c}{2 k}\nabla^2_\bot\right)\es_\pm=
 i g N \hat\sigma_{g e_\pm}.
\label{eq:ShortenedWave}
\end{equation}
The transversal dynamics is described by the last term on the left-hand-side, where $\nabla_\bot = \left[ \partial_{x}, \partial_{y},  0 \right] $ is the transversal vector differential operator.
The dynamics of the atomic system is governed by the Heisenberg-Langevin equations
\begin{equation}
 \dt \hat\sigma_{\mu\nu}=-\gamma_{\mu\nu}\hat\sigma_{\mu\nu}+\frac{i}{\hbar}\left[\hat{H},\hat\sigma_{\mu\nu}\right]+\hat{F}_{\mu\nu},
\label{eq:HeisenbergLangevin}
\end{equation}
where  $\gamma_{\mu\nu}$ are  transversal decay rates and  $\hat{F}_{\mu\nu}$ are $\delta$-correlated Langevin noise operators.

We are interested here in the weak probe field limit, which allows analytical considerations. In this limit we assume
that the Rabi frequency $g$ of the quantum fields $\Es_\pm$ are,
much smaller than $\Omega_\pm$ and the number of photons in the input pulse is much less than the number of atoms.
In this case we can treat the Heisenberg-Langevin equations (\ref{eq:HeisenbergLangevin}) perturbatively in the parameter $g \langle\es_\pm\rangle/\Omega_\pm$.
 We assume that all atoms are in the ground state  in zeroth order, i.e. $\hat\sigma_{gg}=\hat{1}$. In first order
we find
%
\begin{align}
 \dt \hat\sigma_{g e_\pm}\rbt&=-\Gamma_{g e_\pm}\hat\sigma_{g e_\pm}+i g \es_\pm+i \Omega_\mp^* e^{ i\kb_c^\bot \rb}\hat\sigma_{gr}+\hat{F}_{g e_\pm}
\label{eq:Polarizations}\\
\dt \hat\sigma_{g r}\rbt&=-i\hat{\Delta}_{RR}(\rb,t)\hat\sigma_{g r}+i e^{ -i\kb_c^\bot \rb}\sum_{j=\pm}\hat\sigma_{g e_{-j}}\Omega_j
\label{eq:SpinCoherence}
\end{align}
with $\hat{\Delta} _{RR}(\rb)=(N/V_t)\int \hat\sigma_{rr}\rbpt\varepsilon(\rb^\prime-\rb){\rm d}^3 r^\prime$.
Note that the Rydberg state $\ket{r}$ is long-lived and that the decoherence rate $\gamma_{gr}$ is negligibly small \cite{Mohapatra-PRL-2007}.
It would lead to a small loss rate of the considered SL-BEC which we are going to disregard here. Because of the above assumption a Langevin force term $\hat{F}_{gr}$ is not necessary.
For the sake of simplicity we assume that $\gamma_{ge_\pm}=\gamma$, hence $\Gamma_{ge_\pm}=\Gamma=\gamma+i\Delta$.  Within the scope of the publication we restrict ourselves to the case $\Omega=\Omega_+=\Omega_-$ and also assume that we can choose the control field Rabi frequency real \cite{Zimmer-OC-2006}. This is the limit of pure stationary-light. Equations (\ref{eq:Polarizations}), (\ref{eq:SpinCoherence}) together with eq. (\ref{eq:ShortenedWave}) form a self-consistent set of equations and determine the dynamics of the system.
We note that eqs. (\ref{eq:Polarizations}) and (\ref{eq:SpinCoherence}) can be recast in the form
\begin{align}
 \dt \hat{S}\rbt&=-\Gamma \hat{S}+i g \es_S+i \sqrt{2} \Omega(t) e^{ i\kb_c^\bot \rb} \hat\sigma_{gr}+\hat{F}_S,
\label{eq:SumPolarization}\\
 \dt \hat{D}\rbt&=-\Gamma \hat{D}+i g \es_D+\hat{F}_D,
\label{eq:DifferencePolarization}\\
\dt \hat\sigma_{g r}\rbt&=-i\hat{\Delta}_{RR}(\rb,t)\hat\sigma_{g r}+i\sqrt{2}\, \Omega(t) e^{ -i\kb_c^\bot \rb}\, \hat{S},
\label{eq:SpinCoherenceNew}
\end{align}
where we have introduced sum  respectively difference polarization via $\sqrt{2}\hat{S}=\hat\sigma_{g e_+}+\hat\sigma_{g e_-}$ and $\sqrt{2}\hat{D}=\hat\sigma_{g e_+}-\hat\sigma_{g e_-}$. In addition we have defined the new field modes via $\sqrt{2}\es_S=\es_++\es_-$ and  $\sqrt{2}\es_D=\es_+-\es_-$ \cite{Harris-PRL-1994,Zimmer-OC-2006}. Within this new basis the ongoing dynamics becomes more transparent.
\\
For a sufficiently long probe field pulse with characteristic pulse length $T$ we can adiabatically eliminate the difference polarization mode $\hat D$ using eq. (\ref{eq:ShortenedWave})
and eq. (\ref{eq:DifferencePolarization}) if $|\Gamma T|\gg 1$ and find \cite{Zimmer-OC-2006}
\begin{equation}
 \es_D=-L_{\rm abs}\left(1+i\frac{\Delta}{\gamma}\right)\dz \es_S.
\label{eq:DifferenceMode}
\end{equation}
To obtain eq. (\ref{eq:DifferenceMode}) we have furthermore employed that the characteristic length of the probe pulse is much larger then
$\sqrt{\left\vert\ i\Delta/\gamma+1 \right\vert L_{\rm abs}/k}$ where $L_{\rm abs}=\gamma c/g^2 N$ is the absorption length of the medium in absence of EIT.
Using the above result (\ref{eq:DifferenceMode}) and the shortened wave equation for the sum mode (\ref{eq:ShortenedWave}) we find that the sum polarization is given through the sum mode by
\begin{equation}
 \hat{S}=-\frac{i}{g N}\left(\dt-c L_{\rm abs} \left(1+i \frac{\Delta}{\gamma}\right)\dzz-i\frac{c}{2 k}\nabla^2_\bot\right)\es_S.
\label{eq:SumPolarizationFunctionOfSumMode}
\end{equation}
Next we consider eq. (\ref{eq:SumPolarization}) in the limit of an adiabatically slow changing control field.
This limit can be characterized by two features: (a) the Langevin noise operators can be neglected \cite{Zimmer-PRA-2008} and
(b) the coherence between the ground and the Rydberg level is given by
\begin{equation}
\hat\sigma_{gr}=-\frac{g \es_S e^{- i\kb_c^\bot \rb}}{\sqrt{2}\Omega(t) }.
\label{eq:SpinCoherenceAdiabaticLimit}
\end{equation}
If we denote the characteristic time for changes of the control field Rabi frequency by $T$  and if we assume that the density of excitations in the system is small then the adiabaticity conditions
$
\Omega^2/\gamma \sqrt{L_{abs}/L_{pulse}} \gg \frac{1}{T} , \braket{\hat{\Delta}_{RR}\rbt}
$
can easily be satisfied. Here $\braket{\hat{\Delta}_{RR}\rbt}$ is the averaged detuning caused by the dipole-dipole interaction and $L_{pulse}$ is the length of the probe pulse in the medium \cite{Nikoghosyan-PRA-2005}.
Combining eqs. (\ref{eq:SpinCoherenceNew}), (\ref{eq:SumPolarizationFunctionOfSumMode}) and the
result for the spin coherence (\ref{eq:SpinCoherenceAdiabaticLimit}) we arrive at
\begin{align}
 &\left(\dt-c L_{\rm abs} \left(1+i \frac{\Delta}{\gamma}\right)\dzz-i\frac{c}{2 k}\nabla^2_\bot\right)\es_S\rbt\nonumber\\
&=-\frac{g^2 N}{\sqrt{2}\Omega(t)}\left(\dt+i\hat\Delta_{RR}\rbt\right)\left(\frac{\es_S\rbt}{\sqrt{2}\Omega(t)}\right).
\label{eq:SumFieldEquation}
\end{align}
Finally we introduce the dark-state polariton operator of the diamond-type coupling scheme.
We can identify the system's unique dark-state polariton  using  the procedure introduced in \cite{Zimmer-PRA-2008}
\begin{equation}
 \hat\Psi(\rb,t)=\cos\theta \es_S\rbt - \sin\theta\sqrt{N}\hat\sigma_{gr}\rbt e^{i\kb_c^\bot \rb} 
 ,
\label{eq:DarkStatePolariton}
\end{equation}
where the mixing angle $\theta$ is defined via $\tan^2\theta=g^2 N/2\Omega^2$.
 In the adiabatic limit equation eq. (\ref{eq:DarkStatePolariton})  can be simplified to $\es_S\rbt=\cos\theta\hat\Psi\rbt$.
By substituting this expression in eq. (\ref{eq:SumFieldEquation}) we finally arrive at a Schr\"odinger-like propagation equation for the dark-state polariton operator
\begin{align}
i\hbar\dt\hat\Psi\rbt=&\left[
-\frac{\hbar^2}{2 m_\parallel}\dzz-\frac{\hbar^2}{2 m_\bot}\nabla^2_\bot +\hbar\hat\delta_{RR}(\rb)
\right]\hat\Psi\rbt,
\label{eq:DarkPolaritonEquation}
\end{align}
where the dipolar frequency shift in the polariton basis has the form $\hat\delta_{RR}=(N/V_t) \sin^2(\theta)\int \hat\Psi^\dagger(\rb^\prime)\hat\Psi(\rb^\prime)\varepsilon(\rb^\prime-\rb){\rm d}^3 r^\prime$.
In eq. (\ref{eq:DarkPolaritonEquation}) we have introduced the transversal and longitudinal mass using $m_\bot=\hbar k/v_{\rm gr}$, $m_{\parallel }=m_{\bot }\alpha$, where
 $\alpha =\left( 2kL_{abs}\left(\Delta /\gamma -i\right) \right) ^{-1}$.
The group velocity
of the probe field in the EIT medium, which can be controlled by changing the intensity of the coupling
field, is given by $v_{gr}=c\cos ^{2}\theta $.

As noted by Fleischhauer and colleagues Bose-Einstein condensation of SL-DSP can be achieved by raising the
critical temperature of the polariton gas using the variable effective mass of the quasiparticles \cite{Fleischhauer-PRL-2008}.
In contrast to \cite{Fleischhauer-PRL-2008} our scheme allows for direct thermalization via the dipolar interaction of the polaritons.
Well below the critical temperature the system can be described using the order parameter
$\phi(\rb,t)= \sqrt{N/V_t}\braket{\hat\Psi\rbt}$ \cite{Pitaevskii-2003}. Due to this eq. (\ref{eq:DarkPolaritonEquation}) simplifies to
\begin{align}
\left(i\hbar \frac{\partial }{\partial t}\right.%
&\left.+\frac{\hbar^2}{2m_{\parallel }}\frac{\partial^2 }{\partial z^2}+\frac{\hbar^2}{2m_\bot }%
\nabla^2_\bot \right)\phi( \rb,t)  \label{eq:mean_field} \\
&=\hbar \sin^2\theta\int  \left\vert \phi( \rb^\prime,t) \right\vert ^2 \varepsilon ( \rb^\prime -\rb )  \phi( \rb,t)\mathrm{d}^{3} r^\prime, \nonumber
\end{align}
which describes the dynamics of the so called dipolar
Bose-Einstein condensates without contact interaction  \cite{Santos-PRL-2000, Baranov-PR-2008,Lahaye-RPP-2009} .
Note, that the longitudinal and transversal
masses in eq. (\ref{eq:mean_field}) are different. In general the longitudinal mass $m_\parallel$ is imaginary, but, if the single photon detuning is
large compared to the relaxation rate ($\gamma \ll \Delta $) the imaginary
part of $m_\parallel$ can be neglected. Equation
(\ref{eq:mean_field}) is the main result of this manuscript and shows that the dynamics of purely dipolar BECs can be mimicked using stationary light in EIT media.

One of the major problems of dipolar BEC are the instabilities
caused by attractive forces between aligned dipoles. In order to study the
stability properties of (\ref{eq:mean_field}) we calculate the dispersion relation of
weak density perturbations with frequency $\nu $ and wave vector $\bf{q}$
of a homogeneous dipolar SL-BEC with density $n_{dsp}$ \cite{Pitaevskii-2003} and find
\begin{align}
\nu&=\sqrt{\left( \frac{q_{\bot }^{2}}{2m_{\bot }}+\frac{q_{z}^{2}}{%
2m_{\parallel }}\right)
\left( \frac{\hbar ^{2}q_{\bot }^{2}}{2m_{\bot }%
}+\frac{\hbar ^{2}q_{z}^{2}}{2m_{\parallel }}+C_{dd}\left(
3\cos ^{2}\beta -1\right) \right) },  \label{Bogolubov}
\end{align}
where $C_{dd}=8\pi U\wp_r^2 n_{dsp}$. Moreover, $q_\bot=\sqrt{q_{x}^{2}+q_{y}^{2}}$  respectively $q_{z}$ are the transversal and
longitudinal components of $\bf{q}$. The angle between the
excitation wave vector $\bf{q}$ and the direction of dipoles we denote by  $\beta $. In
contrast to the usual dipolar BEC there is no contact interaction term
in the present system \cite{Baranov-PR-2008,Lahaye-RPP-2009}.
In the following we discuss two
cases, defined by the orientation of the considered dipoles with respect to the
direction of propagation of the probe fields, to show the unique properties
of a stationary-light dipolar BEC.

(i) Longitudinal orientation of dipoles
\\
Let us first consider the case when the  dipoles are oriented along the $z$-axis. Then $\cos^{2}\beta
=q_z^{2}/\left( q_\bot^{2}+q_z^{2}\right)
$ and the Bogoliubov spectrum (\ref{Bogolubov}) can be written as
\begin{equation}
\nu =\sqrt{\left( \frac{\tilde{q}^{2}}{2m_{\bot }}\right) \left( \frac{%
\hbar ^{2}\tilde{q}^{2}}{2m_{\bot }}+C_{dd}\left( \frac{2\tilde{q}%
_{z}^{2}\alpha ^{2}-\tilde{q}_{x}^{2}-\tilde{q}_{y}^{2}}{\tilde{q}%
_{z}^{2}\alpha ^{2}+\tilde{q}_{x}^{2}+\tilde{q}_{y}^{2}}\right) \right) ,}
\label{Bogolubov2}
\end{equation}
where ${\bf\tilde{q}}=\left\{ q_{x},q_{y},q_{z}\alpha \right\} $. Since $\alpha
\ll 1$ (the typical value of $\alpha$  is on the order of $10^{-4}$) the cloud is very
stable (unstable) if $C_{dd}<0$
($C_{dd}>0$), the only unstable (stable) region corresponds to the waves
propagating perpendicular (parallel) to $z$ direction due to the small value of $\alpha$. Note that the interaction parameter
$C_{dd}$ can have both positive as well as negative values \cite{Giovanazzi-PRL-2002,Buechler-NP-2007}.

(ii) Transversal orientation of dipoles
\\
For the second case we assume that  the dipoles are oriented along the $y$-axis. Then
the angle between dipoles and wave vector  ${\bf q}$ is defined by $\cos ^{2}\beta
=q_{y}^{2}/\left( q_{\bot }^{2}+q_{z}^{2}\right) $ and for
the excitation spectrum we find%
\begin{equation}
\nu =\sqrt{\left( \frac{\tilde{q}^{2}}{2m_{\bot }}\right) \left( \frac{%
\hbar ^{2}\tilde{q}^{2}}{2m_{\bot }}+C_{dd}\left( \frac{2\tilde{q}%
_{y}^{2}-\tilde{q}_{x}^{2}-\tilde{q}_{z}^{2}\alpha ^{2}}{\tilde{q}%
_{z}^{2}\alpha ^{2}+\tilde{q}_{x}^{2}+\tilde{q}_{y}^{2}}\right) \right). }
\label{Bogolubov3}
\end{equation}
Equation (\ref{Bogolubov3}) indicates that the radial symmetry of interaction is broken and that the dispersion relation becomes
strongly anisotropic in 3D. Namely, for the case of small $\alpha$ and $C_{dd}>0$ the system is stable
(unstable) for the excitations propagating parallel to y-axis (x-axis).

In summary, we have discussed stationary-light in an ensemble of cold Rydberg atoms.
To this end we introduced a diamond-like  light-matter coupling scheme where the upper state is given by a Rydberg state.
We could show that the unique polaritonic quasipartilce behaves in the adiabatic
limit like a Schr\"odinger particle with a purely dipolar inter-particle interaction. Moreover, we could show, by analyzing the Bogoliubov
spectrum of a homogeneous dipolar BEC, that for a special choice of the dipolar interaction parameter the considered dipolar SL-BEC is,
in contrast to usual dipolar BECs, very stable. Depending on the orientation of the Rydberg dipoles used, the SL-BEC shows also a very anisotropic behavior.
The presented scheme opens up new roads between quantum optics and condensed matter physics and puts polaritonic quasiparticles into new light.

Authors are grateful to A. del Campo and T. Pfau for stimulating discussions. F. E. Z. acknowledges the hospitality of the Institute of Theoretical Physics at the University of Ulm. This work was supported by the EU STREP HIP and the Alexander von Humboldt Foundation.
\bibliographystyle{apsrev}
\bibliography{Stationary_Light_in_Rydberg_Vapors_28-02-11-no-title}
\end{document}